\newcommand{\ft}[1]{#1\,fT\,Hz$^{-1/2}$}

\newcommand{\hw}{0.5\linewidth}

\documentclass[12pt]{iopart}
\usepackage{iopams}
\usepackage{harvard,html,url}

\usepackage{graphicx}
\usepackage[caption=false]{subfig}
\usepackage{hyperref} 

\begin{document}

\title[MCG with a modular SERF atomic magnetometer array]{Magnetocardiography
with a modular spin-exchange relaxation free atomic magnetometer array}

\author{R Wyllie$^1$, M Kauer$^1$, G S Smetana$^1$, R T Wakai$^2$ and T G
Walker$^1$}
\address{$^1$ Department of Physics, University of Wisconsin-Madison, 1150 University Ave., Madison, WI  53706}
\address{$^2$ Department of Medical Physics, University of Wisconsin-Madison, 1111 Highland Ave., Madison, WI  53705}

\ead{wyllie@wisc.edu}

\begin{abstract}
We present a portable four-channel atomic magnetometer array operating in the
spin exchange relaxation-free regime. The magnetometer array has several design
features intended to maximize its suitability for biomagnetic measurement,
specifically foetal magnetocardiography, such as a compact modular design and
fibre coupled lasers. The modular design allows the independent positioning and
orientation of each magnetometer. Using this array in a magnetically shielded
room, we acquire adult magnetocadiograms. These measurements were taken with a
\ft{6--11} single-channel baseline sensitivity that is consistent with the
independently measured noise level of the magnetically shielded room.
\end{abstract}


\vspace{2pc}
\submitto{\PMB}
\vspace{2pc}
\maketitle

\newpage


\section{Introduction}
\label{sec:intro}
Human biomagnetism has been an important area of fundamental research and
medicine since the first low noise studies by
\citeasnoun{cohen1970magnetocardiograms}. Magnetic signals can provide
complimentary or unique information in several applications, such as
magnetoencephalography (MEG) \cite{Hamalainen1993} and magnetocardiography (MCG)
\cite{Wakai2000}. For example, high quality foetal MCG (fMCG) signals can direct
{\it in utero} diagnosis and therapy of cardiac arrhythmia such as
supraventricular tachycardia and {\it torsades de pointes}
\cite{Wakai2002,Cuneo2003}.

Typically, these studies use superconducting quantum interference devices
(SQUIDs) for magnetic field detection, usually operated in arrays of matched
gradiometers with 30 or more channels. SQUIDs have achieved a noise limit of
around \ft{1} \cite{Robbes2006}, but they are expensive and require liquid
$^4$He for the low T$_\mathrm{c}$ systems for the high signal-to-noise ratio
(SNR) needed for demanding biomagnetic applications like fMCG \cite{Li2004}.

Atomic magnetometers have recently shown promise for use in biomagnetism. Atomic
vapor magnetometers operating in the spin exchange relaxation free (SERF) regime
\cite{allred2002high} have recently surpassed the sensitivity of SQUID systems,
achieving 160\,aT\,Hz$^{-1/2}$ sensitivities \cite{Dang2010}. Additionally, the
application of microfabrication techniques to the manufacture of mm-scale
alkali-metal vapor cells and integrated fibre-coupled optics \cite{Shah2007}
allows the possibility of large arrays of SERF magnetometers, similar to current
commercial SQUID systems. Besides offering increased sensitivity, atomic
magnetometers are operated at temperatures between 25--180\,C \cite{Budker2007},
replacing expensive cryogenics and maintenance of SQUID systems with simple
resistive electrical heating and passive thermal insulation.

In this paper, we report a four-channel portable atomic magnetometer array
suited for biomagnetic applications. Each magnetometer in the array features a
\ft{6--11} noise floor from 10--100\,Hz and has adjustable channel spacing and
orientation. All channels are self-contained, including local tri-axial nulling
coils, in principle allowing the channels to be placed in an arbitrarily
oriented non-planar geometry. Using this array in a magnetically shielded room
(MSR), we have made high quality adult MCG measurements to demonstrate the
efficacy of our device for use with human subjects.

Previously, adult MEG has been measured using a similar sized fibre
coupled atomic magnetometer operating in a different detection
mode \cite{Johnson2010}. Single channel sensitivity was comparable to ours, but
quadrant photodiode detection was used to construct gradiometer signals, rather
than the use of multiple magnetometers. A similar technique using a photodiode
array was used previously \cite{xia2006meg}, but with a large non-portable
setup. In both cases, the separation between photodiode elements is small
compared the the signal source depth of an MCG measurement. In this regime, the
individual sensors will measure approximately the same signal, and will not
operate as truly multichannel devices on their own.

Adult MCG studies have also been performed. \citeasnoun{bison2009} used an array
of 25 room temperature Cs (non-SERF mode) magnetometers as a 19-channel set of
2nd order gradiometers to map adult MCG in an eddy-current shield. They were able to
suppress the noise by a factor of 1000 using this technique, but were limited to
\ft{300} in this configuration.

Recently, \citeasnoun{knappe2010} used a single channel microfabricated mm-scale
alkali-metal vapor cell to measure adult MCG in an MSR. Their signals compare
well with SQUIDs simultaneously measuring the MCG. The measurements were
performed with a sensitivity \ft{100-200} but in a much quieter shielded room
\cite{Bork2000}. Results from a single magnetometer were presented.

The outline of the rest of the paper is as follows: in \sref{sec:theory}, we
summarize the theory of operation of our SERF magnetometer. \Sref{sec:apparatus}
presents our apparatus and design features. \Sref{sec:methods} details the
operation and characterization procedure of our array. In \sref{sec:mcg}, we
show a sample adult MCG, compare our measured QRS amplitudes from 13 subjects
with previous results in the literature, and demonstrate the acquisition of much
smaller test signals applied with a phantom. \Sref{sec:noise} provides an analysis of the
noise sources limiting our sensitivity. Finally, \sref{sec:conclusion} concludes
with directions for future work towards the goal of obtaining fMCG signals.


\section{Theory}
\label{sec:theory}

The magnetometer presented here operates similarly to previous SERF
magnetometers such as \citeasnoun{allred2002high} and
\citeasnoun{Li2006parametric}. Optical pumping creates a nonzero average
electron spin along the propagation direction of the pumping laser. Orthogonal
magnetic fields cause the population spin to precess at the slowed down Larmor
frequency \cite{Happer1973}. Spin-relaxation collisions and the absorption of
laser photons interrupt the precession. The balance between optical pumping,
spin relaxation, and Larmor precession determines the dynamics of the atomic
polarisation and the performance of the magnetometer. A probe laser detects the
orthogonal components of the electron spin.

\begin{figure}[htp] \centering
	\subfloat[]{
		\includegraphics[width=\hw]{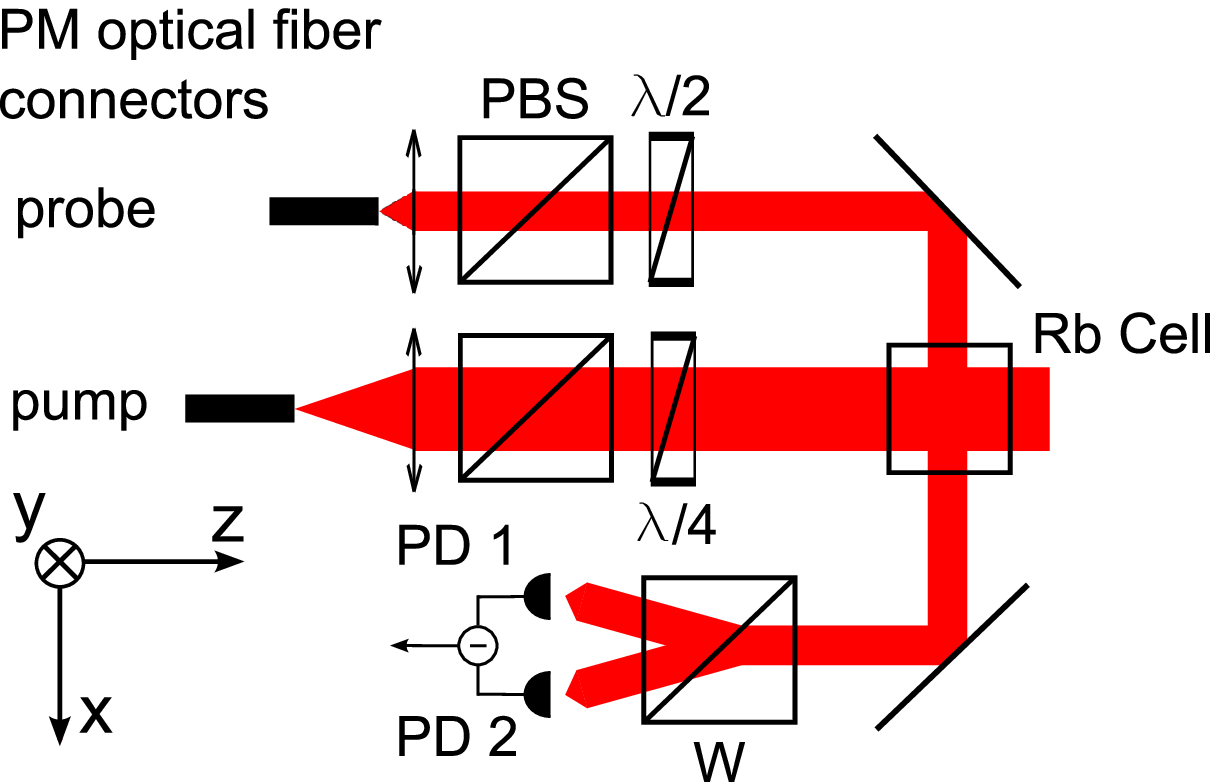}}
		\hspace{0.05\linewidth}
	\subfloat[]{
		\raisebox{+40pt}{\includegraphics[width=0.4\linewidth]{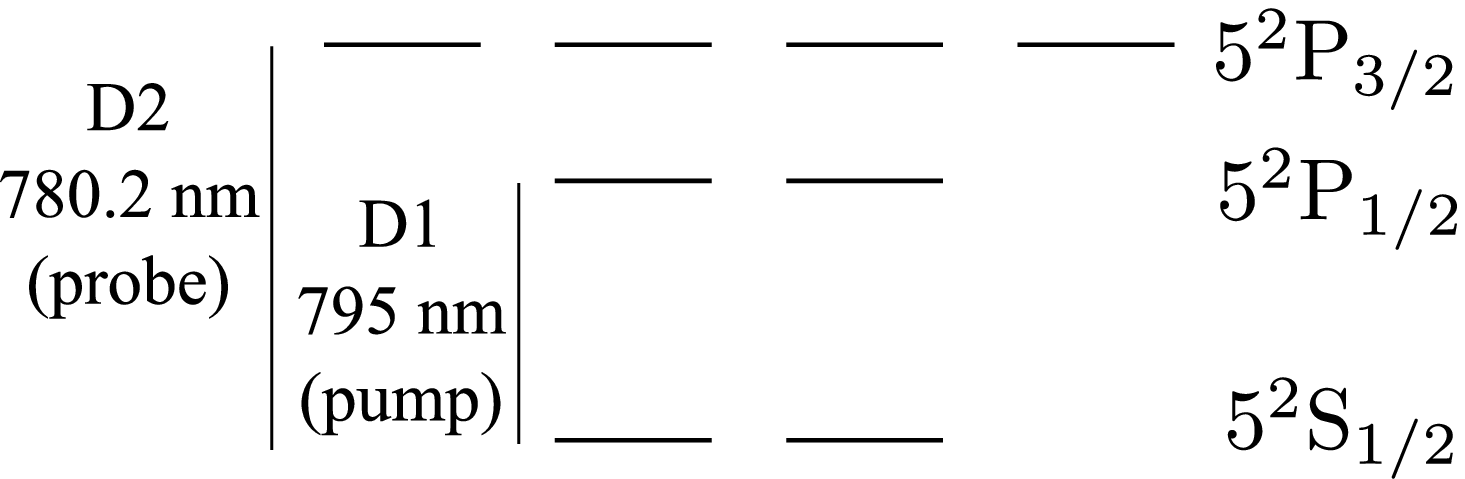}}}
    \caption{(a) Laser beam path schematic. The 795\,nm pump laser is linearly
    polarised by a polarising beam splitter (PBS) and then circularly polarised
    by the quarter waveplate ($\lambda/4$). The 780\,nm probe is detuned from
    the D2 resonance and linearly polarised. Both lasers are conveyed from their
    sources to the optics via polarisation maintaining (PM) optical fibres. The
    half waveplate ($\lambda/2$) is used to balance the photodiode (PD)
    difference signal from the Wollaston polarising beam splitter (W). (b) A
    simplified diagram of the first three fine-structure $^{87}$Rb energy
    levels.}

	\label{fig:laserinfo}
\end{figure}

\Fref{fig:laserinfo}(a) and (b) show the optical setup and $^{87}$Rb energy
level diagram, respectively. A circularly polarised pump beam resonant with the
``D1'' transition optically pumps the atoms. An orthogonal probe beam detuned from the
``D2'' transition detects atomic spin precession due to external fields using
the Faraday effect and a balanced polarimeter. The polarisers at the output of
the optical fibre convert laser polarisation noise into intensity noise, to
which our configuration is much less sensitive.

\subsection{Polarisation equation of motion and steady state solution}

We summarize here a simplified theory of operation of the magnetometer
\cite{Ledbetter2008} to aid interpretation of our results. We use the
simplifying assumption that the absorption linewidth is broad (18\,GHz/(atm
buffer gas)) compared to the $^{87}$Rb ground state hyperfine splitting
(6.8\,GHz), though in our cells the hyperfine splitting and pressure broadening
are nearly equal and the model will be approximate. If the magnetometer is
operated with sufficiently high Rb density and sufficiently low total magnetic
field, spin-exchange collisions preserve the coherence of the transverse spin
\cite{allred2002high} and three relaxation processes dominate the polarization
relaxation rate $\Gamma$,
\begin{equation}\label{eq:gamma}
\Gamma = \Gamma_{\mathrm{sd}} + \Gamma_{\mathrm{pr}} + R,
\end{equation}
where $R$ is the optical pumping rate (the rate that an unpolarised atom would
absorb pump photons), $\Gamma_\mathrm{sd}$ is the spin-destruction rate from collisions
and $\Gamma_\mathrm{pr}$ is the absorption rate of probe laser photons.

Assuming the Rb atoms are in spin-temperature equilibrium
\cite{Anderson1960} and neglecting diffusion, the polarisation has a
steady-state solution for static magnetic fields \cite{Ledbetter2008}
\begin{equation}\label{eq:sspx}
P_x=P_z\left(\frac{\Gamma\Omega_y+\Omega_x\Omega_z}{\Gamma^2+\Omega_z^2}\right)
\approx P_z \frac{\Omega_y}{\Gamma},
\end{equation}
where $\bOmega = \gamma {\bf B}$ is the Larmor frequency from a magnetic field
{\bf B} and $\gamma=g_\mathrm{s} \mu_\mathrm{B}/ \hbar$. ${\bf P}$ is the Rb
polarisation in the coordinate system of \fref{fig:laserinfo}(a). The
approximations on the right hand side of \eref{eq:sspx} hold
for small magnetic fields such that $|\Omega| \ll \Gamma$. At typical $\Gamma
\approx 2\pi\times 1000$/s, this limits B to $\ll 30$\,nT.
%
%
 \subsection{Signal size and sensitivity}

The Rb polarisation modifies the susceptibility of the atomic vapor, which
changes the index of refraction for $\sigma _+$ and $\sigma _-$ circularly
polarised light. For linearly polarised light, this causes a rotation of the
polarisation angle. In our case, the probe beam is linearly polarised in the
$\hat{y}-\hat{z}$ plane and is rotated in this plane by \cite{Ledbetter2008}

\begin{equation}\label{eq:rotation}
\phi \approx \frac{1}{4}r_\mathrm{e} f c n l
\frac{(\nu-\nu_0)}{(\nu-\nu_0)^2+(\Delta\nu/2)^2}P_x,
\end{equation}
where $r_\mathrm{e}$ is the classical electron radius, $f$ is the transition
oscillator strength, $c$ is the speed of light, $n$ is the Rb atomic number
density, $l$ is the optical path length, $\nu-\nu_0$ is the probe laser detuning
from the D2 resonance, and $\Delta\nu$ is the pressure broadened absorption
linewidth of the transition. For reference, using our experimental parameters
(see \sref{sec:noise}), \eref{eq:rotation} evaluates to $\phi/B_y\approx
0.1$\,$\mu$Rad/fT. For perspective, the differential photocurrent is $\Delta
I=2P_0\epsilon\phi$, with probe laser power $P_0$ and a photodiode power to
current conversion $\epsilon$, is around 250\,pA/fT.
%
%
\section{Apparatus and design considerations}
\label{sec:apparatus}
\subsection{Apparatus description}
\begin{figure}[htp]\centering
	\subfloat[]{
		\raisebox{+20pt}{\includegraphics[width=.6\linewidth]{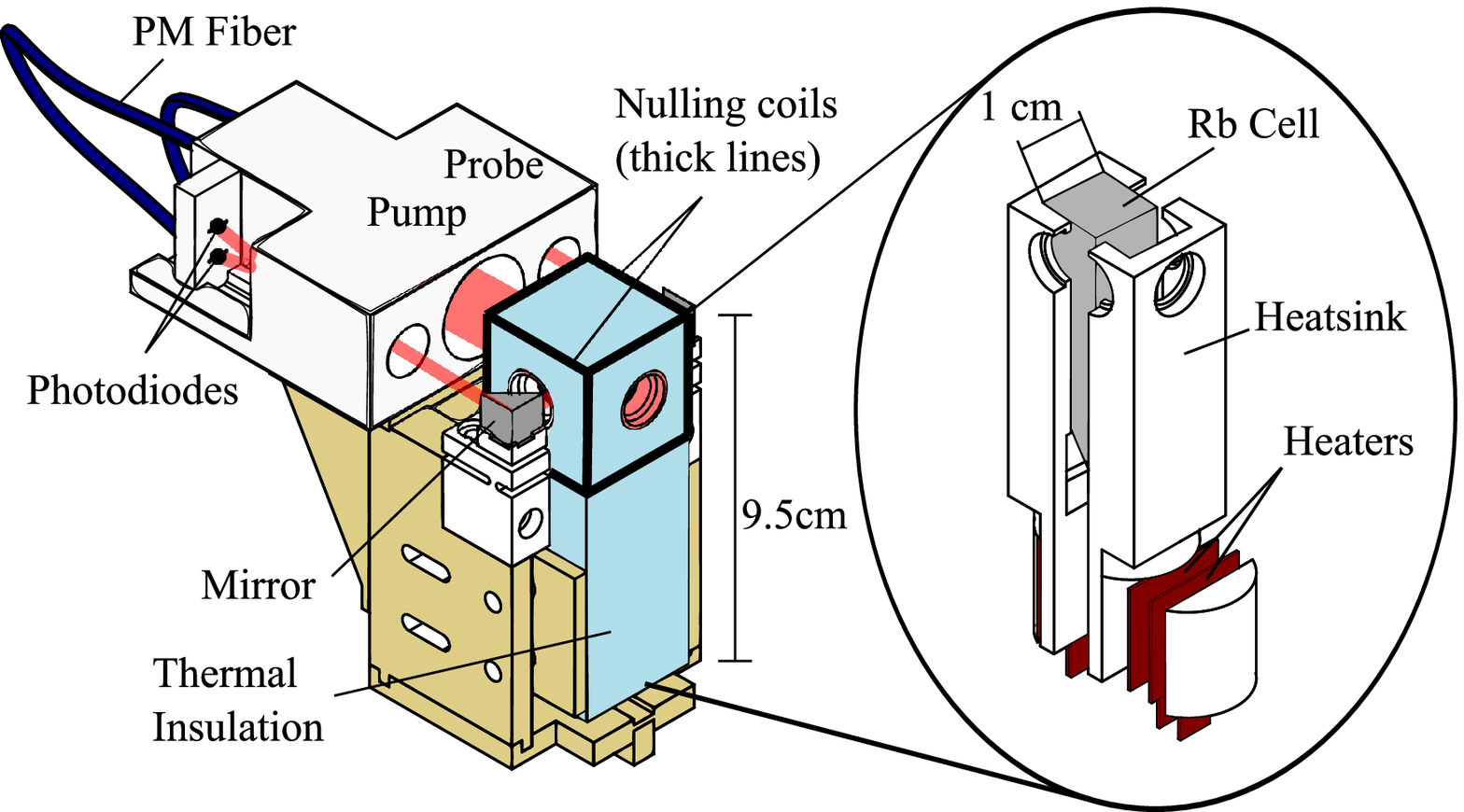}}}
		\hfill
	\subfloat[]{
		\includegraphics[width=0.35\linewidth]{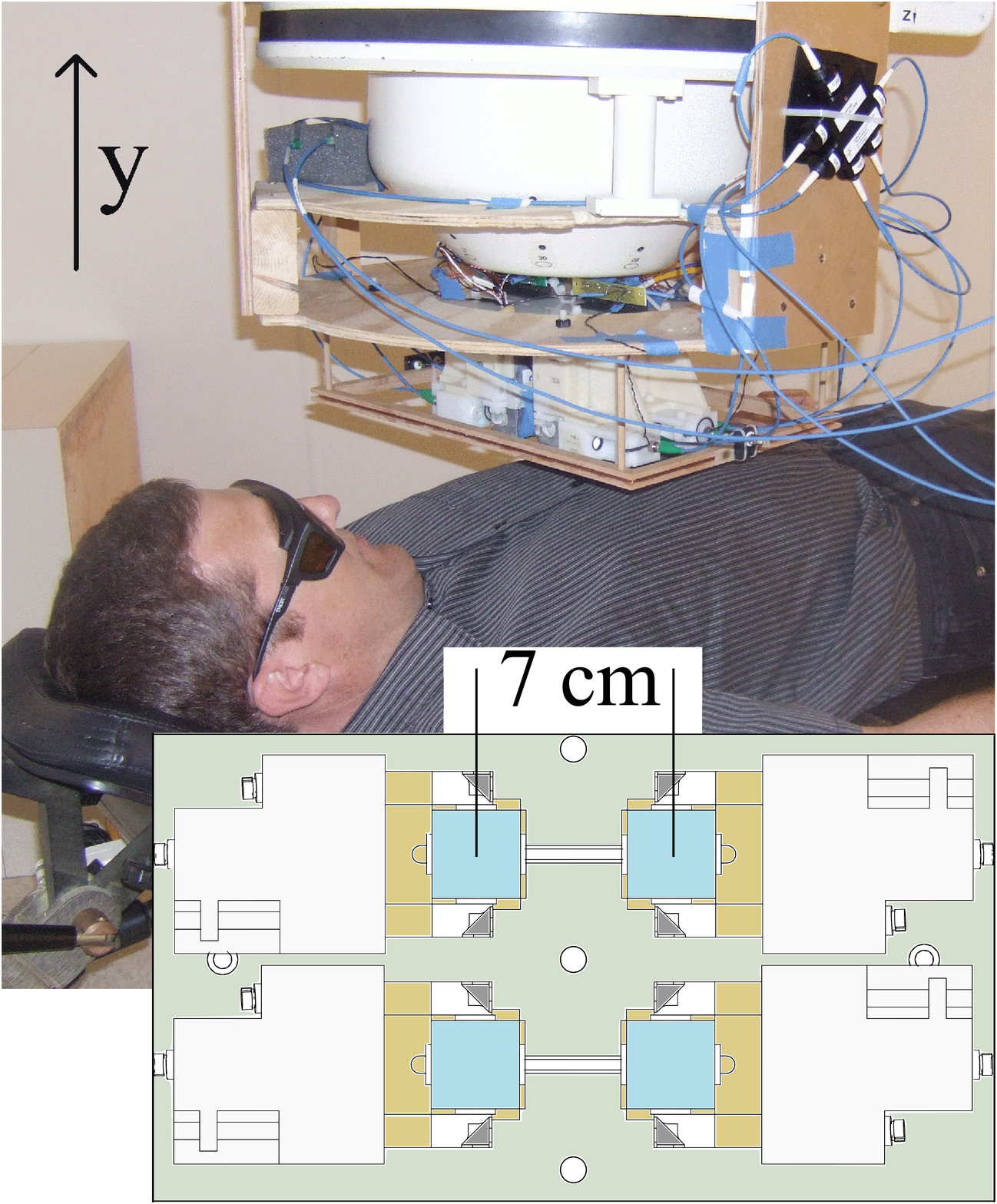}}
\caption{(a) Single magnetometer unit (left) with view inside the thermal
insulation (right). Square tri-axial nulling coils are wrapped around the
insulation. (b)(bottom insert) Four channel magnetometer array, with 7\,cm
channel spacing. This array is mounted on an existing SQUID gantry top-down, as
in (b)(top). In human subject trials, the subject lays on a bed and the array is
centered over the subject's heart by equalizing the signal in all four channels
in real time. The $\hat{x}$, $\hat{y}$, and $\hat{z}$ axes in the text refer to
magnetometer body axes. Here, all $\hat{y}$ are aligned and correspond to the
sensitive direction for all magnetometers.}
\label{fig:array}
\end{figure}
Design choices in this atomic magnetometer are meant to minimize noise
consistent with maximizing suitability for biomagnetic measurements.
\Fref{fig:array}(a) shows a single magnetometer unit. It consists of three fixed
optics tubes with all the optics from the schematic in \fref{fig:laserinfo}(a),
a Pyrex $^{87}$Rb vapor cell (Triad Technologies) press-fit inside a
boron-nitride heatsink thermally contacted to resistive-film heaters (Minco) and
a thermistor, all surrounded by thermal insulation (Aerogel). A plastic scaffold
clamps the heatsink, optics tube, and two custom-made plastic mirror mounts in
place. Single-turn triaxial coils are wrapped around the thermal insulation to
locally null the magnetic fields in the detection volume and provide calibration
and control fields. Polarisation maintaining fibres connect the optics tube to
the pump and probe lasers. The pump laser is a distributed feedback type with
output power 80\,mW (Eagleyard Photonics), detuned $|\Delta
\nu_\mathrm{pump}|\le 20$\,GHz from the D1 resonance. The probe laser is a
20\,mW tunable external cavity diode laser (Sacher Lasertechnik Group), detuned
$\sim$0.2\,nm from the D2 resonance. The two lasers are each split equally into
four separate fibres with an integrated polarisation maintaining splitter
(Oz-Optics). The result is a set of individual magnetometer channels that are
unconstrained with respect to one another, due to local fibre coupled optics and
individual magnetic field control at each cell.

The lasers are mounted on a portable breadboard which, along with all the coil
control electronics and monitoring equipment, fit on a single cart. A PXI based
FPGA data acquisition unit (National Instruments) is used to apply calibration
fields and digitize magnetometer signals. The PXI crate also houses power supply
cards to control the vapor cell heaters. This modular design ensures the entire
apparatus is easily transported between the MSR and our testing lab in a
different building.

Material selection was aimed at minimizing Johnson noise caused by the use of
electrical conductors. Each magnetometer unit shown in \fref{fig:array}(a)
contains only a few conductive pieces, namely the heaters, thermistor leads,
optical fibre terminations, and the photodiode leads.

The vapor cell is 1\,x\,1\,x\,5 cm Pyrex rectangular container with isotopically
pure $^{87}$Rb and buffer gas composed of 50\,Torr nitrogen to prevent radiation
trapping \cite{Happer1972} and 760\,Torr He at the time of manufacture.
Subsequent measurement showed slow loss of buffer gas pressure indicating He
loss consistent with diffusion through the Pyrex cell walls
\cite{Walters1970}. We estimate the data presented here was taken with a
residual He pressure of $\sim$200\,Torr. The cells were operated at a
temperature of 140--180\,C. Each magnetometer requires 4\,W of heater power in
thermal equilibrium, corresponding to electrical currents near 0.5\,A, large
enough to produce potentially disruptive magnetic fields and gradients. We manage these
using matched heater pairs with oppositely aligned magnetic fields and a boron
nitride heatsink to spatially separate the heaters from the detection volume of
the vapor cell. Furthermore, the heater currents were modulated at 100\,kHz so
these magnetic fields and their gradient fields would average to zero on the time
scale of MCG measurements. There is no detectable difference between the
magnetic noise level with the heaters on or off.

\Fref{fig:array}(b) shows a photograph of the array mounted to an existing
gantry for a previously installed commercial SQUID system. Human subjects lie on
a bed under the array. The array can be translated, rotated, and tilted using
the gantry controls. For convenience in these initial tests, our measurements
were made with planar square array and channel spacing of 7\,cm. The minimum
planar array spacing for all four elements is 4.5\,cm. More magnetometers can be
added, with 7\,cm spacing in the vertical direction of
\fref{fig:array}(b)(bottom), while the optics tubes require the spacing in the
horizontal direction to be $>$10\,cm. In our setup, the four elements operate
independently and could be tilted or translated with respect to one another. It
should be straightforward to implement non-planar geometries.

\subsection{Modularity}\label{sec:modularity}

\begin{figure}[htp] \centering
	\subfloat[]{
		\includegraphics[width=0.45\linewidth]{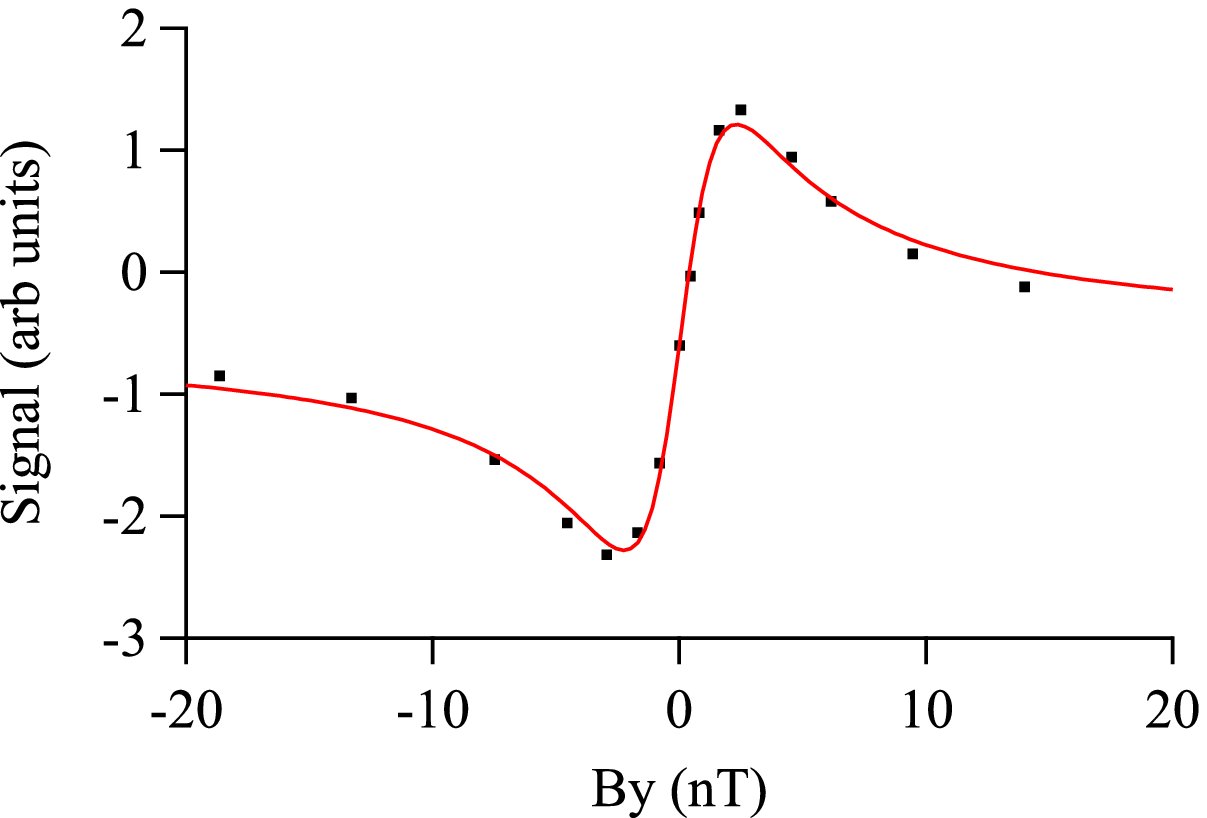}}
	\hfill
	\subfloat[]{
		\includegraphics[width=0.45\linewidth]{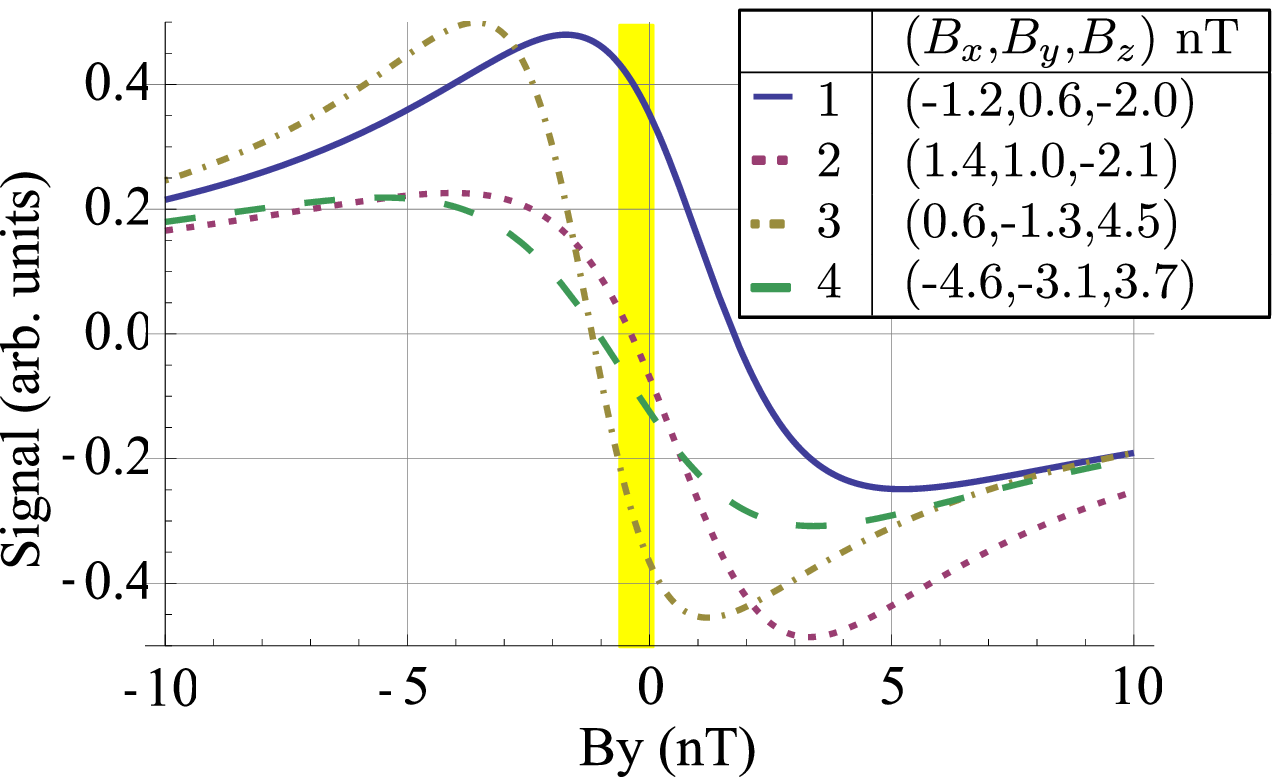}}
\caption{(a) Magnetometer signal as a function of applied field (dots) and a fit
to an amplitude scaled version of \eref{eq:sspx}. (b) A simulation of $P_x$ for
multiple magnetometers in the presence of the measured magnetic field gradients
from the zero values at each channel. With only a single set of tri-axial coils,
these are the minimum gradients in the shielded room, and represent the
best-case operating point for the four magnetometers. In this case, a compromise
value for $B_y$ must be chosen, reducing the usable operating range. This
illustrates the need for local control over the magnetic fields at each cell.}
	\label{fig:dispersion}
\end{figure}

DC fields in the MSR, 10--50\,nT, decrease the fundamental sensitivity according
to \eref{eq:sspx} and require nulling. \Fref{fig:dispersion}(a) shows a sweep of
$B_y$ and a fit with $R=400\approx \Gamma$ to \eref{eq:sspx} with
$\Omega_x,\Omega_z=0$. $B_y$ must be nulled to better than $|B_y|\le 2$\,nT to
remain in the sensitive linear region of the response.

If the MSR magnetic fields were sufficiently uniform, we could use a single set
of tri-axial coils with side length $a$ and array channel baseline spacing $d$
such that $a \gg d$ to simultaneously cancel the magnetic fields at all
magnetometers. Having tried this, we have found two problems with this approach
that resulted in the choice of small individual cancellation coils for each magnetometer.

First, residual MSR magnetic field gradients are too large on the scale of the
spacing of our magnetometers to allow a uniform nulling field to be used.
\Fref{fig:dispersion}(b) shows the response predicted by \eref{eq:sspx} for four
magnetometers in the presence of measured MSR magnetic field gradients and $R$
from the fit in \fref{fig:dispersion}(a). $B_x$, $B_y$, and $B_z$ are the
differences from the measured zero values at each magnetometer channel. As shown
in \fref{fig:dispersion}(b), these gradients cause a decrease in both the single
channel sensitivity and the effective multichannel $B_y$ operating range
(highlighted in yellow), which has been decreased by a factor of 4 to $|B_y| \le
0.5$\,nT.

Second, and most importantly, the pumping laser can cause a light shift (or AC
Stark shift) due to the pump laser interacting differently with different Zeeman sublevels through the
Rb vector polarisability \cite{savukov2005effects,Budker2007}. This causes an
effective magnetic field \cite{Appelt1999}
\begin{equation}\label{eq:starkshift}
{\bf \Omega}_\mathrm{LS}\approx R\Delta\hat{s},
\end{equation}
where $\Delta$ is the ratio of the laser detuning from atomic line centre to
absorption linewidth, and $\hat{s}$ is the photon helicity, $\pm1$ along the
pump propagation direction for circularly polarised light.

At high optical depths corresponding to high temperatures, the pump beam is
absorbed before polarising the entire cell length unless it is detuned
$\Delta$\,$\sim$\,1 from resonance, resulting in a nonzero ${\bf
\Omega}_\mathrm{LS}$. The magnitude $|{\bf \Omega}_\mathrm{LS}|$ is roughly the
same from cell to cell, but the direction with respect to the null coils depends on the
orientation of the magnetometer and the direction of $\hat{s}$. For our laser
parameters, we expect the pump light to cause effective fields of magnitude
$\sim$\,50\,nT. Orienting all the magnetometers to have the same light shift
with respect to the external coils mitigates the problem, though the nonuniform light
shift within each cell due to the spatial intensity profile of the laser results
in a broadening of the magnetometer dispersion \cite{Appelt1999}.

The light shift presents a problem for the prospects of differing individual
channel orientation and tilt, or nonplanar arrays. There are also practical
problems associated with large coils, such as limited magnetometer adjustment and
complications with calibration fields when the magnetometers are not aligned
with the coil axis. These problems are all solved by using individual sets of
coils for each magnetometer.


\section{Methods}
\label{sec:methods}

\subsection{Residual Field Nulling}

We find the DC null fields for a single magnetometer as follows
\cite{seltzer2004unshielded,Li2006}:
\begin{itemize}
  \item sweep $B_y$ and set the operating point in the centre of the
  dispersive curve, where the magnetometer is most sensitive.
  \item apply a 50\,pT oscillating field $B_0\sin(\omega t)\hat{x}$ at
  10--30\,Hz and adjust $B_z$ until there is no response to $B_0$, ensuring
  $B_y$ remains at its most sensitive operating point.
  \item apply $B_0\sin(\omega t)\hat{z}$ and adjust $B_x$ to null the
  response, again ensuring $B_y$ is adjusted to maintain maximum
  sensitivity.
  \item repeat until residual response to $B_x$ and $B_z$ is minimized.
\end{itemize}
For the four-magnetometer array, we repeat this procedure for each individual
unit. Iterating the procedure two times eliminates any offsets of the local
field at the first magnetometer caused by adjusting the fields in the later
ones to below the adjustment procedure sensitivity.

There is no cross-talk between our four channels because the magnetometers are
fundamentally passive devices. Non-uniform light shifts from the Gaussian pump
laser intensity profile cause non-uniform effective magnetic fields as in
\eref{eq:starkshift}. Due to the residual $\Omega_z$ term in \eref{eq:sspx},
each magnetometer retains some sensitivity to $B_x$. This is typically a
correction of $\le 10\%$ and its effect is ignored in what follows.

As discussed in \sref{sec:modularity}, the magnetometers must be within
approximately 1\,nT of thier null point in order to maintain uncompromised
sensitivity. Background field variations in the shielded room are generally
smaller by an order of magnitude or more on the timescales of of a 30\,s MCG
measurement. Long-term variations can take the magnetometers outside of the
optimal measurement range and are easily compensated between measurements.

\subsection{Magnetometer sensitivity and noise floor characterization}
\label{sec:MSR_noise}

\begin{figure}[htp]\centering
 \includegraphics[width=.9\linewidth]{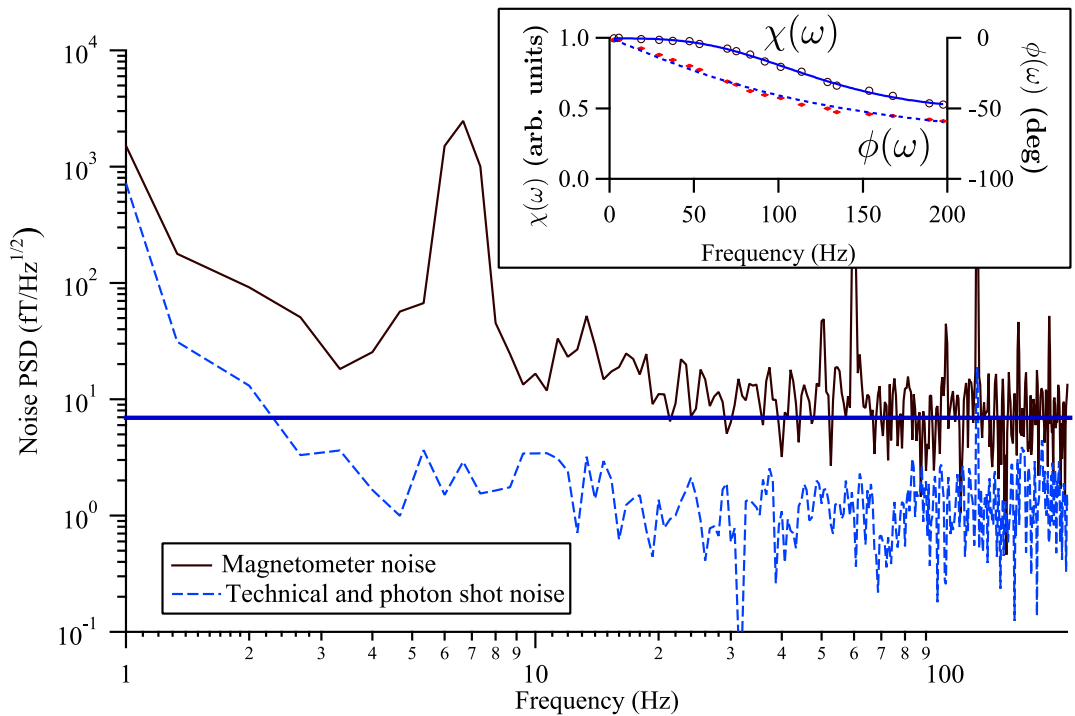}
\caption[Single channel Calibration and Noise PSD]{Measured magnetic noise
(solid black) of a single characteristic magnetometer channel on a magnetically
quiet day in the MSR. (Inset) Amplitude response $\chi(\omega)$ (open circles)
and phase response $\phi(\omega)$ (dots) of a single magnetometer to a
calibrated applied signal. The fits to $\chi(\omega)$ and $\phi(\omega)$ are
used to calibrate the all noise PSD and time-series measurements presented here.
The horizontal line  is at \ft{6}, the expected magnetic noise level based on
\tref{tab:noise}. The no-pump effective noise level (dashed blue) is \ft{1--2},
consistent with our estimates for photon shot noise.}
	\label{fig:cal}
\end{figure}

Once the DC null fields have been determined, each magnetometer is
simultaneously characterized by applying a set of 10--200\,pT magnetic fields at
twenty different frequencies ranging from 2--200\,Hz, as in
\fref{fig:cal}(inset), providing a frequency dependent calibration for each
magnetometer signal $S$, such that $S=\chi
(\omega)e^{\mathrm{i}\phi(\omega)}B_y$. We then take a short ($\sim$6\,s) noise
measurement detecting the residual response of the magnetometer to ambient
fields and other noise sources. Each magnetometer is sampled at 20\,kHz with a
two-pole lowpass filter at 10\,kHz, using a 16-bit digitizer. The
power spectral density (PSD) of this signal is used to characterize the noise
floor as
\begin{equation}\label{eq:psd} 
\frac{\delta B}{\sqrt{\mathrm{Hz}}}= \frac{\sqrt{\mathrm{PSD}}}{\chi(\omega)},
\end{equation}
producing the measurement in \fref{fig:cal}. We note that while the MSR floor is
constant over time, sinusoidal residual magnetic fields often dominate the
spectrum. \Fref{fig:cal} shows a measurement with a relatively quiet background,
displaying only a few such peaks near at 6.7, 50.5, 60, 101, and 120\,Hz. We
suspect these peaks come from nearby air handling mechanics and powerline noise.
Though this makes real time analysis of the magnetometer signals difficult, they
tend not to interfere with MCG measurement because they are narrowband and
easily identified.

To obtain an estimate of the non-magnetic noise level of the magnetometers, we
block the pump beam and measure the PSD once again. In this case, the device
should not be sensitive to magnetic fields. This is non-magnetic noise that is
transformed into an effective noise floor through the same calibration
$\chi(\omega)$ used in \eref{eq:psd}. The residual noise is from photon shot
noise, spin-projection noise, and non-magnetic technical noise.
\Fref{fig:cal}(inset) shows an example of the single magnetometer calibration
parameters $\chi(\omega)$ and $\phi(\omega)$. The blue dashed curve of
\Fref{fig:cal} shows the non-magnetic noise level. All magnetometers operated
with a baseline sensitivity $\le$\,\ft{11} and a probe-noise limited sensitivity
at least a factor of two below the measured magnetic field sensitivity.


\section{MCG}
\label{sec:mcg}

\begin{figure}[htp]
\centering
\includegraphics[height=3.71in]{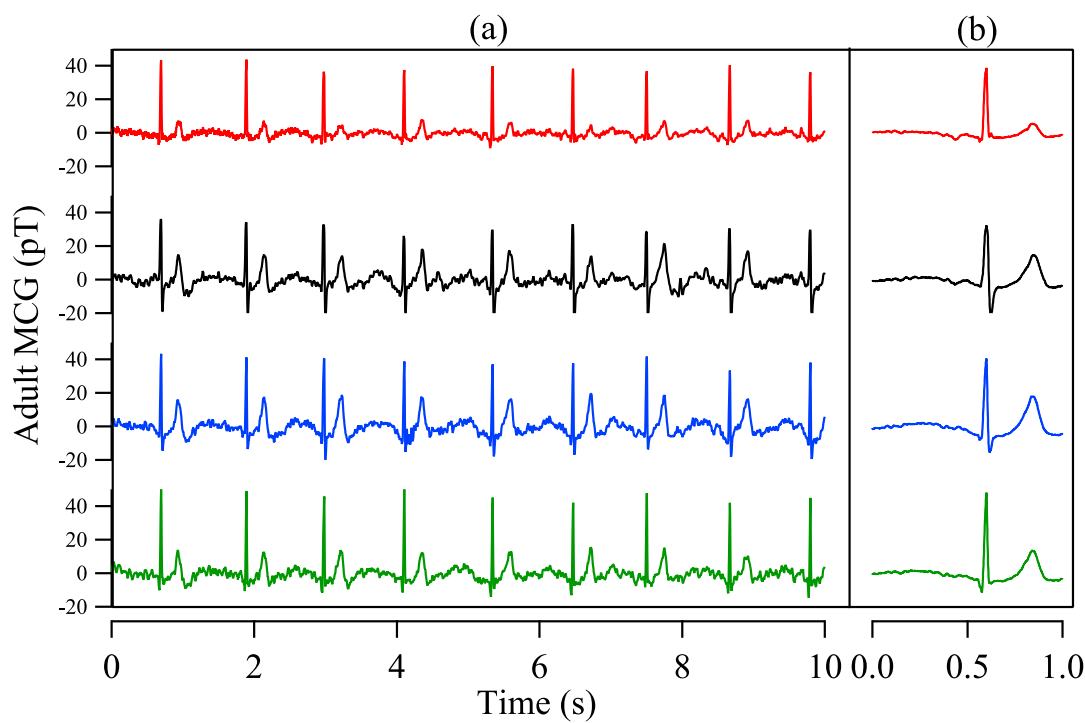}
\caption{(a) Four-channel adult MCG with large signal interferences subtracted.
(b) Shows the same set of MCG after averaging each channel.}
\label{fig:4mcg}
\end{figure}

We obtained MCG from 13 adult human subjects. The experimental protocol was
approved by the institutional human subjects committee prior to commencement of
the study and informed consent was obtained from all subjects. 

Each measurement consists of a 30\,s MCG recording from each magnetometer
sampled with the same parameters described above. For data analysis, the
magnetic field in the time-domain is obtained by a deconvolution of the
magnetometer response (measured in the calibration) from the measured signal.
The signal is then down-sampled to 1024\,Hz. Low frequency drift is removed
using a wavelet detrending algorithm.

\begin{figure}[htp]
\centering
\includegraphics[width=1\linewidth]{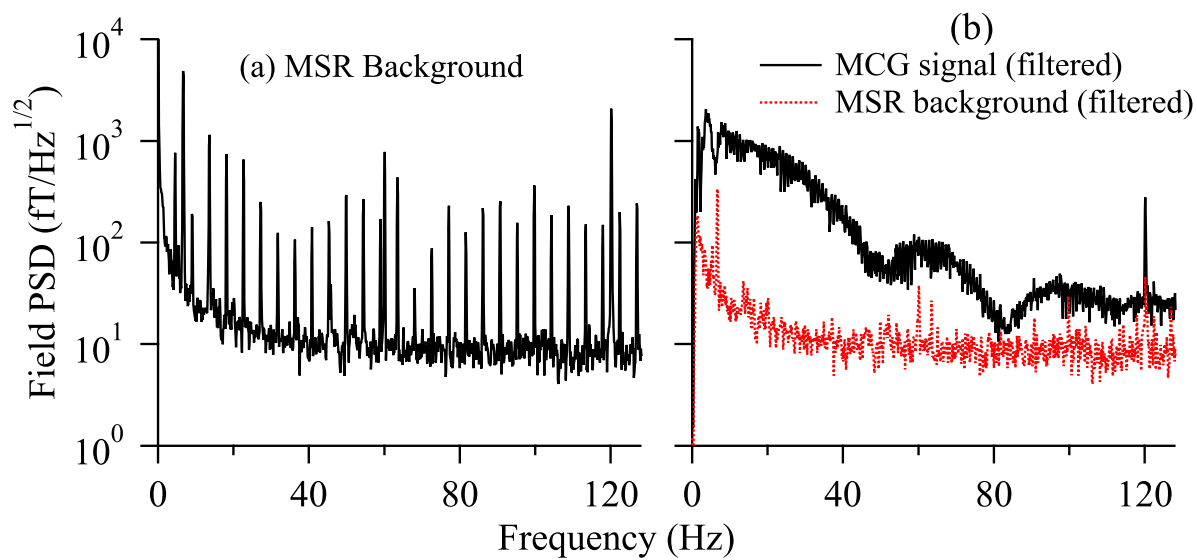}
\caption{(a) MSR noise background for the MCG data. The peaks are actual
magnetic artifacts, likely from building air-handling units, and comparison
with \fref{fig:cal}, obtained on a different day, illustrates the day-to-day
variation in the MSR background fields. (b) MCG signal (black) and MSR noise
background (red dots) after the signal processing described in the text. The
non-biomagnetic peaks in (a) are easily removed.}
\label{fig:psdref}
\end{figure}

\Fref{fig:4mcg}(a) shows a processed segment of 30\,s MCG recordings from the
four channels. \Fref{fig:psdref}(a) shows a reference measurement taken just
after the data presented in \fref{fig:4mcg}, but without a subject. This
measurement was taken on a different day than \fref{fig:cal} and there are
clearly more peaks, mostly at 4.55\,Hz and its harmonics, along with a prominent
6.7\,Hz peak and powerline harmonics. To obtain the traces in
\fref{fig:4mcg}(a), the magnitude and phase of these non-biomagnetic peaks is
determined from the reference, and then a matching sinusoid is subtracted from
the signal channels, resulting in the PSDs shown in \fref{fig:psdref}(b).
\Fref{fig:4mcg}(b) shows the averaged traces. The P, QRS, and T components are
well resolved. We emphasize these are magnetometer signals (rather than
gradiometer signals), and contain any unfiltered background magnetic noise in
the MSR as well as the MCG signal.

We estimate the minimum distance to the skin for each magnetometer of around
1\,cm. \citeasnoun{knappe2010} reported their sensor head volume was located
5\,mm from the skin or 5\,cm from the heart, whereas the SQUIDS used for
comparison were 7.5\,cm from the heart. With our channel spacing (7\,cm), we
estimate each magnetometer is 7.6\,cm from the heart when the array is centered
above it.

\begin{table}
\caption{\label{tab:qrs}QRS peak amplitudes of this work. QRS dipole moments
are estimated to compare with previous works. The distance range quoted here is
for the range of individual magnetometer distances to the heart.}

\footnotesize
\hfill
\begin{indented}
\item[]\begin{tabular}{@{}l c c c}
\br
&~\cite{Nousiainen1994a}~&~\cite{knappe2010}~&~This work\\
\mr
$r$ (cm)~&~8--10.5~&~5~&~5.5--10.8\\
\mr
$B_\mathrm{QRS}$ (pT)~&~(not reported)~&~100--150~&~$40\pm16$\\
\mr
$m$ ($\mathrm{\mu}$Am$^2$)~&~0.76
$\pm0.36$~&~0.4--0.6$^\mathrm{a}$~&~$0.60\pm0.48^\mathrm{a}$\\
\br
\end{tabular}
\item[] $^\mathrm{a}$ Estimated from $B_\mathrm{QRS}$ and $r$.
\end{indented}
\end{table}

A statistical summary of the QRS peaks of 13 subjects is presented in
\tref{tab:qrs}, along with an estimation of the QRS dipole amplitude $m$. We use
the approximation $B_\mathrm{QRS}\sim m/r^{2.6}$ \cite{Nousiainen1994a} to
account for distributed source effects. To calculate $m$, we require the
distance from the heart to each magnetometer. We minimize the standard deviation
of $m$ across the four magnetometers as a function of the heart position with
respect to the center of the magnetometer array, which can then be used to
calculate the dipole amplitude for each patient. The result from this work in
\tref{tab:qrs} is a weighted mean of the 13 measured subjects.

\begin{figure}[htp] \centering
\includegraphics[width=\linewidth]{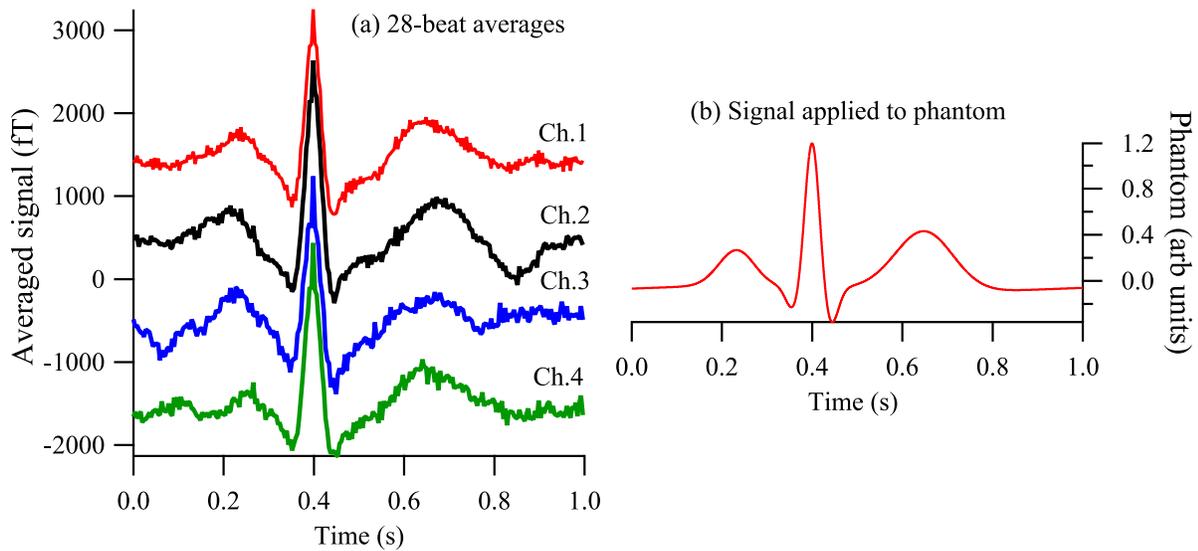} \caption[Phantom MCG]{(a)
28-beat average response from the phantom driven with the waveform shown in (b),
with magnetic field offsets applied to aid visualization. Before averaging, the
signals are processed in the same way as the MCG above, by removing only narrow
frequency components.}
	\label{fig:phantom}
\end{figure}

In addition to adult MCG measurements, we also simulated MCG using a head
phantom (Biomagnetic Technologies), which is a spherical shell filled with
saline solution and five different current dipoles that can be driven
individually. The head was placed as close as possible to the array centre and
aligned so that the detected signal from driving a dipole was maximized. We
drove the phantom with a test waveform \cite{Goldberger2000} shown in
\fref{fig:phantom}(b).

We used the calibrated magnetometers to adjust the QRS peak height driven by an
amplitude scaled version of the test waveform. \Fref{fig:phantom}(a)
shows 28-beat averages for all four magnetometer channels, after processing the
signals in the same way as the MCGs. The detected QRS amplitude is about 2\,pT,
and the averages for each channel faithfully represent the field expected from
the driving waveform.


\section{Noise sources and characterization}
\label{sec:noise}

There are several different sources of unwanted signal in our magnetometer.
These fall into three general categories. One is technical noise, including
electrical pickup noise or probe movement on the detector photodiodes. The
second is magnetic signal other than the desired biomagnetic signals, and the
third noise category is fundamental quantum noise associated with the magnetic
measurement. The last two terms are the dominant noise sources and we estimate
the noise floor in \tref{tab:noise}.

\begin{table}
\caption{\label{tab:noise}Magnetometer Noise Budget. Two results are
quoted for the photon shot noise. The basic model is \eref{eq:psntheory} where
$P_z$, $\Gamma_\mathrm{pr}$, and $\Gamma$ are all averages over the laser
interaction volume. In the detailed model, these quantities are calculated at
each position within the cell and then integrated to recover the photon shot
noise floor.}

\begin{indented}
\lineup
\item[]\begin{tabular}{@{}l l}
\br 
Noise Source & Level
(\ft{})\\
\mr
MSR Johnson noise~&~5\\
Electrical current noise in nulling coils~&~3.9\\
$^{87}$Rb Johnson noise~&~0.9\\
Photon shot noise (basic model)~&~0.3\\
Photon shot noise (detailed model)~&~2\\
Spin-projection noise~&~0.03\\
\mr
Total~&~6.2\\
\br
\end{tabular}
\end{indented}
\end{table}


\subsection{Magnetic noise}

Magnetic noise generally includes any magnetic source that falls within the
frequency band of interest and limits the achievable SNR of the desired signal.
Clear examples can be seen in \fref{fig:psdref}(a).

The single channel magnetometer noise floor in \fref{fig:cal}(b) and
\fref{fig:psdref} is limited by Johnson noise in nearby electrical conductors.
These conductors carry thermal currents which generate white-noise profile
magnetic fields. The most important noise sources in our case are likely the
mu-metal walls of the MSR and $^{87}$Rb thin-film remaining on the Pyrex cell
walls. We estimate these contributions from the calculations of
\citeasnoun{lee2008calculation}. The 2\,m cube shielded room has a noise floor
of around \ft{5}. We roughly model the $^{87}$Rb film as a cylinder with a 1\,cm
radius and 10\,nm thickness, about 5\,mm from the detection volume, which would
have a \ft{0.9} noise level.

We also measured the electrical current noise in the circuitry used to drive our
nulling coils using a transimpedance amplifier. This contributes \ft{3.9} to the
magnetic noise level.

\subsection{Photon shot noise}\label{sec:psn}

Though most SERF magnetometers are limited by shield Johnson noise, there are
two fundamental quantum noise sources which limit the maximum theoretical magnetic
field resolution: spin-projection noise and photon shot noise. We briefly discuss
the results in \citeasnoun{Ledbetter2008} as they pertain here to photon shot
noise. We estimate spin-projection noise to be a negligible contribution to the
overall noise budget.

Photon shot noise occurs because the polarisation of the probe laser is
determined by counting the difference in the number of photons arriving at the
two different photodiodes in \fref{fig:laserinfo}(a). The resulting uncertainty
in the magnetic field is

\begin{equation}\label{eq:psntheory}
\delta\Omega_y=\frac{2 \Gamma}{P_z \sqrt{\Gamma_\mathrm{pr}
\mathrm{OD}_0 n V t}},
\end{equation}
where $\mathrm{OD}_0=2 r_\mathrm{e} f c n l / \Delta\gamma$ is the probe optical
depth on resonance.

The parameters used in our calculations were $R=2575$/s,
$\Gamma_\mathrm{sd}=60$/s, $\Gamma_\mathrm{pr}=155$/s, $n=7.8\times
10^{13}$/cm$^3$, pump waist $w_{0,\mathrm{pu}}=2.6$\,mm, and probe waist
$w_{0,\mathrm{pr}}=1.1$\,mm, and interaction volume $V=\pi w_{0,\mathrm{pr}}^2
2 w_{0,\mathrm{pu}}$. \Eref{eq:psntheory} estimates a photon shot noise level of
\ft{0.3}.

We have found using this calculation with spatially averaged laser parameters
estimates a lower photon shot noise limited sensitivity than we actually achieve
in our experiments. A more detailed \cite{Walker1997} model approximating the
pump and probe laser propagation through the cell and their Gaussian spatial
profiles, gives a more realistic \ft{2}.

For comparison, an experimental measure of the photon shot noise can be obtained
using the measured differential photocurrent per unit field $\Delta I_p/\Delta
B$ and an effective photocurrent shot noise where $I_p$ is the photocurrent from
a single photodiode. The photon quantization noise is $\delta
I_p/\sqrt{\textnormal{Hz}}=\sqrt{2eI_p}$. The effective magnetic field noise is
then

\begin{equation}\label{eq:psnexp}
\frac{\delta B}{\sqrt{\textnormal{Hz}}}=\frac{\delta
I_p/\sqrt{\textnormal{Hz}}}{\Delta I_p/\Delta B}.
\end{equation}

For a typical measured ${\Delta I_p/\Delta B}=8$ $\mathrm{\mu A}/\mathrm{nT}$
operating with all four channels and a total probe power of 1.67\,mW,
$\delta B/ \sqrt{\textnormal{Hz}}=$ \ft{1.4} from \eref{eq:psnexp}, in
reasonable agreement with the more detailed calculation.


\section{Conclusions and future work}
\label{sec:conclusion}

We have presented a modular four-channel atomic magnetometer array operating in
the SERF regime. The adjustable 7\,cm channel spacing used for these
measurements is much larger than previously presented biomagnetic measurements using SERF
magnetometers. The single channel sensitivity ranged from \ft{6--11}, dominated
by magnetic noise from the MSR. The modularity of the magnetometers, in
principle, allows flexible channel positioning in an array, despite large light
shifts in different directions in the room reference frame.

Our sensitivity is limited by real magnetic noise. In this case, parametric
modulation \cite{Li2006parametric} can allow simultaneous measurement of $B_x$
and $B_y$, at the cost of a small (1/2\,--\,1/4) decrease in signal size. Future
work will aim at using parametric modulation to measure two components of the
magnetic field with each modular unit.

Detection of fMCG, an application which requires high sensitivity but a modest
number of channels, should be possible with our array. However, the number of
channels is insufficient to allow the effective use of spatial filters, such as
beamformers and independent component analysis, which are widely used to remove
maternal interference and to improve the signal-to-noise ratio of fMCG
recordings.  With a modest reduction in size, the design can be extended to
higher channel counts.  This will be essential for constructing dense arrays
suitable for MEG and adult MCG.

\ack
This work is supported by NIH grant number 5R01HD057965-02.

\newpage
\section*{References}
\bibliographystyle{jphysicsB}
\bibliography{library}

\end{document}